# Dark Web Activity Classification Using Deep Learning

Ali Fayzi[a*], Mohammad Fayzi[b], Kourosh Dadashtabar Ahmadi[c]

[a] Machine Learning Engineer at AIFA (Artificial Intelligence Group of FANAP CO), a.fayzi@fanap.ir
[b] Department of information management, Management Faculty, Tehran Branch, Islamic Azad University, Tehran, Iran, MohammadFayzi@outlook.com
[c] Department of Information Technology Auckland Institute of Studies, kourosha@ais.ac.nz

**ABSTRACT**

**In contemporary times, people rely heavily on the internet and search engines to obtain information, either directly or indirectly. However, the information accessible to users constitutes merely 4% of the overall information present on the internet, which is commonly known as the surface web. The remaining information that eludes search engines is called the deep web. The deep web encompasses deliberately hidden information, such as personal email accounts, social media accounts, online banking accounts, and other confidential data. The deep web contains several critical applications, including databases of universities, banks, and civil records, which are off-limits and illegal to access. The dark web is a subset of the deep web that provides an ideal platform for criminals and smugglers to engage in illicit activities, such as drug trafficking, weapon smuggling, selling stolen bank cards, and money laundering. In this article, we propose a search engine that employs deep learning to detect the titles of activities on the dark web. We focus on five categories of activities, including drug trading, weapon trading, selling stolen bank cards, selling fake IDs, and selling illegal currencies. Our aim is to extract relevant images from websites with a ".onion" extension and identify the titles of websites without images by extracting keywords from the text of the pages. Furthermore, we introduce a dataset of images called "Darkoob", which we have gathered and used to evaluate our proposed method. Our experimental results demonstrate that the proposed method achieves an accuracy rate of 94% on the test dataset.**

## 1. Introduction

Dark web is a subsection of the vast global network, which requires special software to access it. Once accessed, websites and other services available on the dark web can be accessed via a browser similar to regular web browsing. Websites that are completely hidden on the dark web exist, to the extent that even search engines cannot crawl their pages, and the only way to access them is through the use of a global address.

There are specific online stores on the dark web, known as "Darknet Markets," which mainly sell illegal products such as drugs, firearms, and stolen bank cards, with payments usually made using virtual currencies such as Bitcoin. The majority of cybercrime activities taking place on the dark web are illegal, such as the buying and selling of drugs and weapons, the buying and selling of sexual slaves, and even online physical and sexual abuse of humans, all without leaving any trace of individuals.

However, the functionality of the dark web is not limited to the above-mentioned fields. Journalists and political activists in countries under severe internet censorship can also benefit from this method without leaving any trace of themselves. The dark web was initially used by the United States Navy, but eventually became a haven for criminals due to its privacy and anonymity features.

Another feature of the dark web is that communication between users and servers is established anonymously, meaning that neither the user nor the server has access to the real address of the other. Initially, the purpose of these networks was to establish untraceable communication with military, but over time it has become a general tool for those who need anonymous communication.

The dark web is often considered synonymous with the deep web, but in reality, the two networks have differences, and both do not operate in the same fields and areas. The deep web refers to all websites that cannot be accessed through search engines and are not indexed by them. Therefore, the dark web can be considered a subset of the deep web, but it is different in terms of the techniques and services offered on this network. These services and products are not available to all internet users, and only a specific group of users who are interested in accessing illegal services and products can access them, such as websites selling drugs, illegal betting, child abuse websites, and websites selling military weapons. In short, websites containing illegal services and products on the deep web can be referred to as the dark web.

Based on the above explanations, the web can be divided into three categories, as shown in figure (1): a small part that is accessible to ordinary individuals, which is called the public web (clear web), the majority of which is the deep web, which is not crawled by search engines. Finally, the dark web is a small subset of this space that mainly deals with illegal activities.

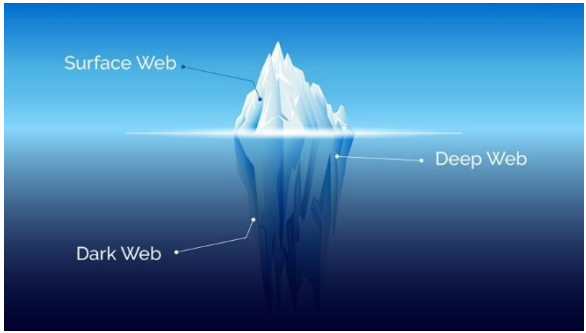

Figure. 1. Division of the dark web space into three subcategories: open web, deep web, and dark web

OSINT tools are used to extract information from the web, including the dark web. Discovery, collection, and monitoring are key processes in OSINT. Methods such as advanced searches, social network monitoring, and accessing restricted content are used to retrieve desired information from the surface web. However, the dark web's unique features and traffic distribution require traditional OSINT methods to be compatible with the dark web's laws and a specific technical configuration to access dark web data. OSINT uses publicly available information for analysis and interpretation.

## 2. Related works

Collecting and analyzing data from the dark web requires intelligent tools that deep web networks have been able to demonstrate their effectiveness in performing such tasks. In this article, we examine various methods that first explore the structure of the dark web, and then employ deep neural networks to analyze the information obtained from this space.

According to the article [1], the criminal network nature that uses the dark web is rarely understood and rarely studied, primarily due to data scarcity. Darknet markets (DNMs), online sales platforms on the dark web, are among the main components of the underground economy. Due to their anonymity and increased access to these platforms, they are rich sources of cyber threats such as hacking tools, data breaches, and personal account information. With the increasing number of products offered in DNM, researchers have begun developing machine learning-based threat identification approaches. The main challenge in adopting such an approach is that it usually requires manually labeled training data, which is expensive and impractical.

This article proposes a new semi-supervised labeling approach to penetrate unlabeled data based on linguistic and structural features of DNMs using transfer learning. Empirical results show that the proposed approach leads to an increase of nearly 3-5% in the measured classification performance with an F1 score while increasing both precision and recall. To improve identification performance, the article [2] adopts a long short-term memory (LSTM) as a deep learning structure in the proposed labeling approach. The results have been evaluated using a large set of 79,000 product listings obtained from the most popular DNMs. The proposed method in article [2] has better performance in identifying threats compared to advanced methods and is considered an important step towards reducing the cost of human supervision in achieving automatic threat detection in cyber threat intelligence organizations.

One of the tasks of law enforcement agencies is to find evidence based on criminal activity in the dark web. However, manually visiting thousands of domains to find visually informative data containing illegal activities requires significant time and resources. Additionally, the background of images can create challenges during classification. To solve this problem, in [3], automatic classification of dark web images has been done using a point-wise attention filtering strategy, which filters out irrelevant image features at the pixel level that do not belong to the object of interest by combining saliency maps with visual bag-of-words (BoVW). SAKF has evaluated on a custom Tor image dataset against CNN features where MobileNet v1 [4] and Resnet50 with dense SIFT descriptors have achieved accuracy of 87.98% and better performance than all other approaches.

Complex anonymity mechanisms and difficult tracking in the dark web provide a refuge for illegal activities. Illegal content on the dark web is diverse and often updated. Traditional dark web classification uses large-scale web pages for supervised training. However, the difficulty of collecting illegal content in the dark web and the time required for manual web page labeling have turned into current research challenges. In [5], a proposed method can effectively classify illegal content on the dark web using a combination of deep learning and transfer learning techniques. The proposed method uses a pre-trained deep learning model for feature extraction and classification. The results show that the proposed method outperforms traditional classification methods, achieving an accuracy of 92.5%.

One of the main features of TOR, the most well-known dark web from the deep web, is its high level of anonymity provided to its users. Advanced research has shown that the TOR network displays a wide range of legal and illegal services and activities, including file sharing, ransomware, and counterfeit goods. In [6], a framework is presented for identifying some of the previously mentioned services in TOR through their image content. In this article, the DUSI service images are introduced to make them publicly available (an image-based dataset containing pictures of active TOR domains from six different service categories). Using DUSI, two pipelines are evaluated based on perceptual hashes and visual bag-of-words (BoVW). In the first pipeline, the hash code is calculated and then compared with the previous saved hashes, which indicate each service, using Hamming distance. In the second pipeline, multiple SVM models trained with BoVW feature vectors encoded with SIFT and Edge-SIFT descriptors are independently and combinatively trained. In the classification stage, each new image is encoded using the relevant vocabulary and classified by the trained models. The highest accuracy achieved was 99.38%, and the fact that it does not require a trained model, transforms perceptual hashing into the recommended approach for identifying TOR services through a snapshot of their homepages.

Regarding deep learning and artificial neural networks-based detection of activity titles in the dark web space, various research studies have been conducted in [2], [3], [5],[6]. In [2], the most important point emphasized is the provision of the required dataset for training deep learning networks. In [3], automatic classification of dark web images is performed using a point-wise attention filtering strategy that combines saliency maps with visual bag-of-words to filter out irrelevant image features at the pixel level. In [5], a combination of deep learning and transfer learning techniques is proposed to effectively classify illegal content on the dark web, achieving an accuracy of 92.5%. Finally, in [6], a framework is presented for identifying TOR services through their image content, achieving high accuracy without requiring a trained model. These studies demonstrate the potential of deep learning and neural networks for analyzing and detecting activity titles in the dark web space, which can improve the efficiency of law enforcement agencies in identifying criminal activities while reducing the need for manual supervision and classification.

In summary, the dark web presents significant challenges for data analysis due to its anonymity and the illegal activities that take place within it. However, recent research has shown that deep learning techniques, combined with transfer learning and attention filtering strategies, can effectively analyze and classify data from the dark web, including identifying cyber threats and illegal content. These approaches have the potential to improve the efficiency and effectiveness of law enforcement agencies in identifying criminal activities in the dark web while reducing the need for manual supervision and classification.

.

## 3. Image CLASSIFICATION

One of the main areas of artificial intelligence (AI) is computer vision, which involves various applications such as image classification. For instance, for object detection algorithms to recognize an object, it is necessary to determine which class or category the identified objects belong to.

As shown in the figure(2), an example of the ImageNet dataset [7] is presented, which is a very large dataset used for object classification operations.

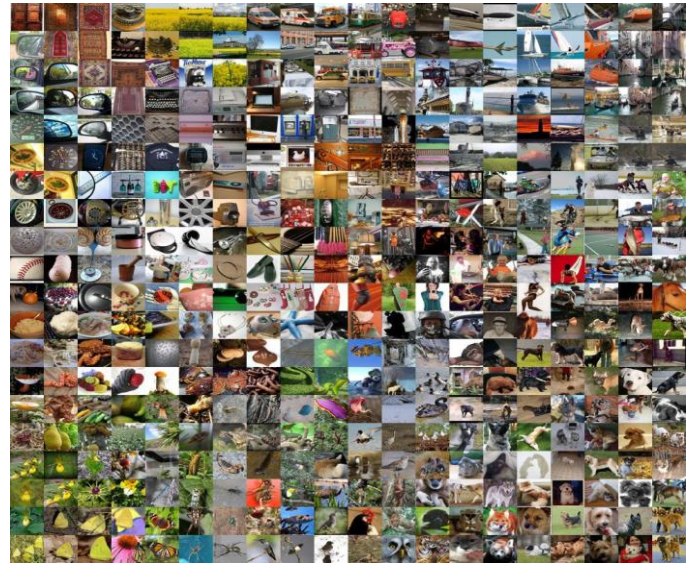

Figure. 2. sample images from the ImageNet dataset [7]

Image classification presents numerous challenges and difficulties for algorithms. As shown in Figure(3), one of these challenges is intra-class variations, where a single class can have significant visual differences, such as different types of handguns or various appearances, sizes, and shapes of identity documents.

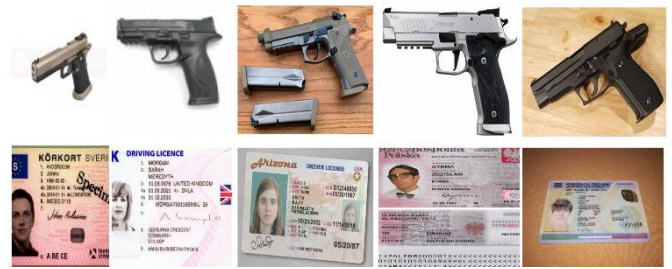

Figure. 3. An example of intra-class changes in weapons and identification cards

Another challenge in image classification is the imaging conditions and lighting environment. Many images have different shapes from various angles, and the position of the light source also significantly affects the captured image. Other issues include blurry images, small object sizes, shape changes such as humans in sitting or standing positions, the presence of shadows, objects with similar colors to the background, and many other cases that are not addressed in this study.

## 4. proposed method

Given the challenges of detecting activity titles in the dark web space, the proposed method addresses the low-quality images available in the dark web, as well as the variations in the angles of the images and the possibility of parts of the images being obscured for various reasons, such as obstruction. The proposed method aims to overcome these challenges through advanced techniques in computer vision, such as perceptual hashing, visual bag-of-words, and deep learning, to achieve high accuracy in identifying activity titles in the dark web space. The proposed method can significantly improve the efficiency of law enforcement agencies in identifying criminal activities in the dark web while reducing the need for manual supervision and classification.

### 4.1. Dataset

According to research conducted in the field of detecting activity titles in the dark web, there is a dataset called TOIC [10], but unfortunately, access to it was not provided due to various reasons. Therefore, we decided to collect a dataset for detecting activity titles, which resulted in creating a dataset with five different categories, as follows:

1. Buying and selling drugs
2. Buying and selling weapons
3. Buying and selling bank cards
4. Buying and selling identity documents
5. Buying and selling illegal currencies

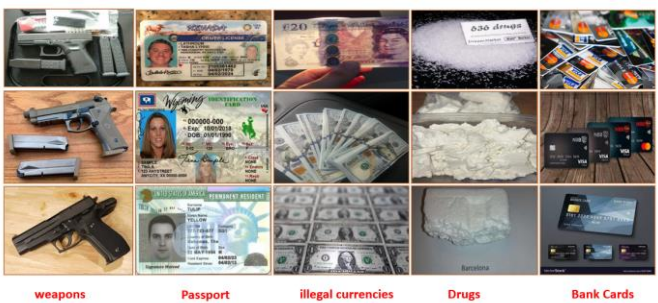

Figure. 4.   An example of the darkoob dataset

The dataset is named " darkoob " by the authors, and in the following sections, we introduce the different parts of the dataset and the various operations performed on it to increase the number of data. The collected dataset after cleaning operations contains 1355 data, and the details are presented in the table 1.

Table 1.   darkoob dataset

| Category Name | Count |
|---|---|
| Drugs | 252 |
| weapons | 319 |
| illegal currencies | 229 |
| Passport | 229 |
| Bank Cards | 295 |

The TOIC dataset contains 698 images in 5 categories similar to the categories in the darkoob dataset. The dataset collected by us is almost twice the size of the TOIC dataset. The darkoob dataset was collected to address the limitations of the TOIC dataset and to improve the accuracy of detecting criminal activities in the dark web. The additional data in the darkoob dataset can enhance the performance of machine learning algorithms and deep learning models in detecting activity titles and help law enforcement agencies to combat criminal activities more effectively.

### 4.2. Data collection process

In order to obtain images from the dark web, we first need a list of dark web websites. For this purpose, a Python script was created that collects a list of internet addresses of various dark web websites with the ".onion" extension in 5 categories (drugs, weapons, illegal currencies, identity documents, and bank cards). After collecting the internet addresses of the dark web websites, we need to connect to the dark web to download the images available on these websites. We use the TOR network for this purpose, and finally, using another Python script and the TOR network, we can download the images available on the dark web websites and categorize them into different categories and perform data cleaning operations. The total number of downloaded images for the 5 categories is 20908, which reduces to 1355 after removing duplicate and unusable images. The overall performance of the dataset collection system can be seen in Figure(5). The darkoob dataset was collected to improve the accuracy of detecting criminal activities in the dark web and to provide law enforcement agencies with a more comprehensive dataset for combating criminal activities.

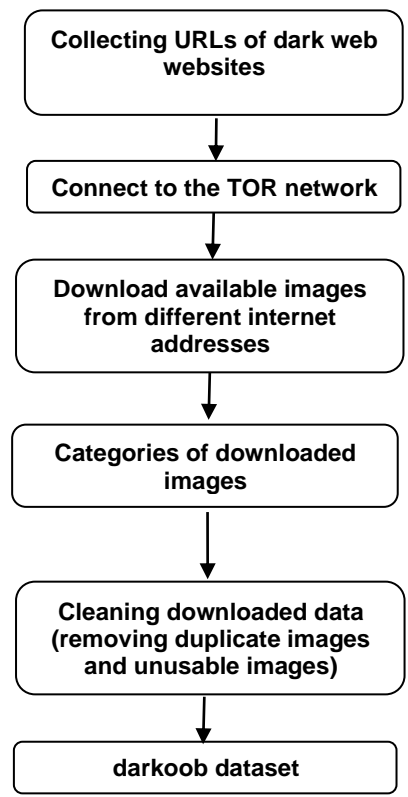

Figure. 5.   The process of collecting the darkoob dataset

### 4.3. Data Augmentation

A convolutional neural network (CNN) that can classify objects even when they are oriented in different directions will have the feature of invariance. Specifically, a CNN can be invariant to translation, viewpoint, size, and/or brightness intensity (or a combination of these).

This is essentially a data augmentation hypothesis. In the real world, we may have a database whose images are taken under restricted conditions, but our target program may exist under different conditions such as orientation, position, scale, brightness, etc. These situations are taken into account by training the neural network with artificially modified data.

In this research, several simple data augmentation methods were performed on the introduced dataset, and other researchers who want to work on these datasets can use other data augmentation methods on this dataset. In the figure(6), several examples of the performed data augmentation operations will be reported.

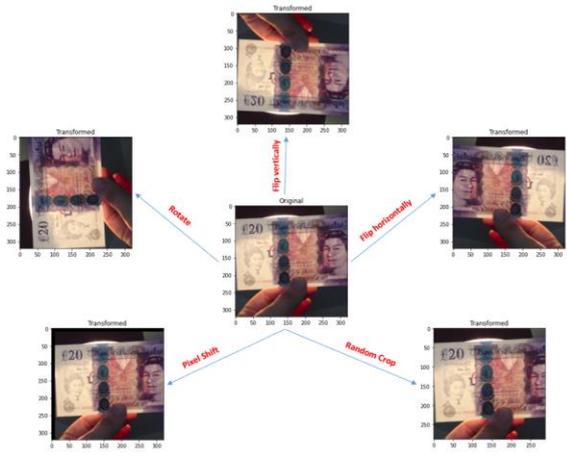

Figure. 6. set of data augmentation operations

After the data increase operation on the darkoob data set, the number of data has increased to 6381, the table of which is as follows.

Table 2. Darkoob dataset after data augmentation

| Category Name | Count |
|---|---|
| Drugs | 1098 |
| weapons | 2871 |
| illegal currencies | 945 |
| Passport | 531 |
| Bank Cards | 936 |

In the image below, you can see randomly selected samples from the darkoob dataset, which shows different images from different categories.

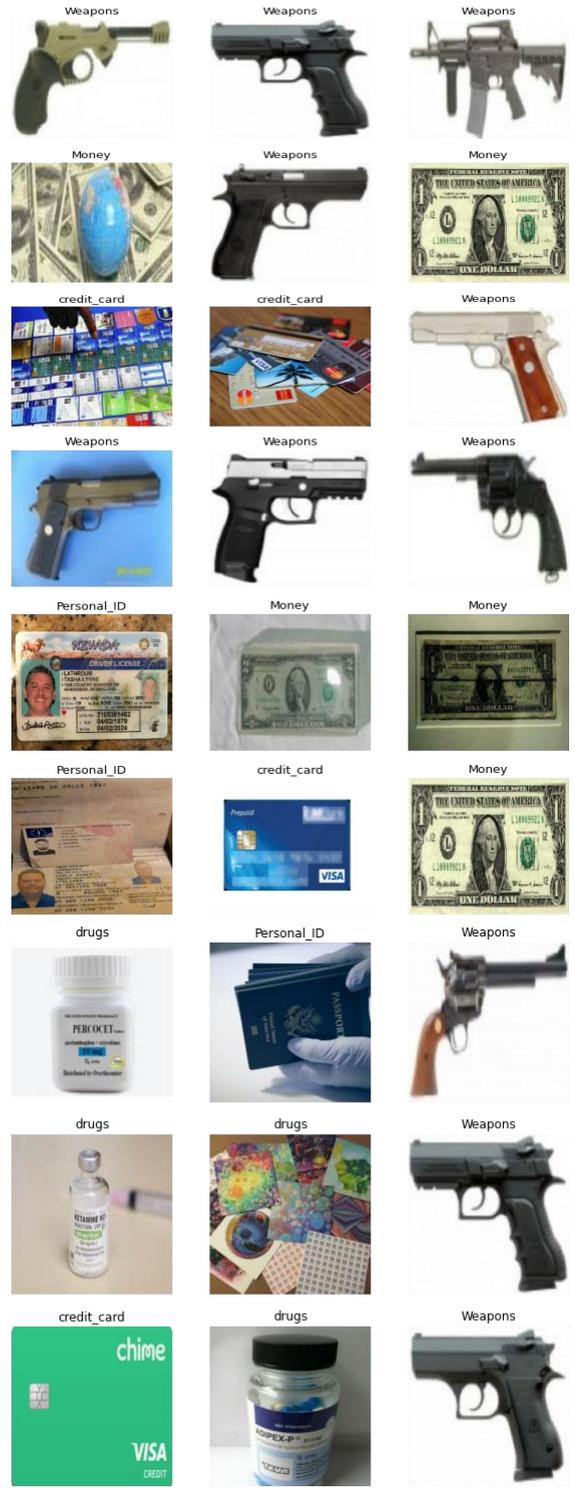

Figure. 7. Random samples from the darkoob data set

### 4.4. Deep Learning

Due to the sensitivity of detecting activity titles in the dark web, we need to execute this system using intelligent systems. In this research, deep neural networks and CNN algorithms are used to perform image recognition operations on images extracted from the dark web.

Furthermore, a new dataset collected from the dark web is introduced in this research, which focuses on 5 categories of different activities in the dark web. The darkoob dataset was created to address the limitations of the TOIC dataset and provide a more comprehensive dataset for detecting criminal activities in the dark web. The dataset includes images of illegal drugs, weapons, bank cards, identity documents, and illegal currencies, which can be used to train machine learning algorithms and deep learning models to improve the accuracy of detecting criminal activities. The darkoob dataset can be a valuable resource for law enforcement agencies and researchers in the field of cybersecurity and can contribute to the development of more effective solutions for combating criminal activities in the dark web.

### 4.5. Convolutional Neural Network

Convolutional neural networks (CNNs) were first introduced by Fukushima in 1988 [12]. However, due to computational costs, they did not gain much popularity. Later in the 1990s, LeCun and colleagues used CNNs for handwritten digit classification [13]. The biggest breakthrough came in 2012 with AlexNet [8], which won the ImageNet Large Scale Visual Recognition Challenge. Since then, CNNs have been widely used, particularly for visual images. As their name suggests, CNNs derive their power from convolution operations. In the convolution operation, depending on the scale, it considers a pixel and its surrounding pixels. This allows for the continuous extraction of local features.

CNNs have a different architecture compared to other neural networks because their inputs are images instead of numbers. Additionally, there are no neurons in CNNs, but rather convolutional filters exist in their hidden layers. These filters are applied to the images as inputs or feature maps for hidden layers, and the filter weights are adjusted during the training stage. These convolutional filters act like neurons in ANN, applied to images or feature maps, and their results are fed into an activation function. A general architecture of CNNs is shown in the figure.

CNNs have shown great success in various image recognition tasks, including object detection, face recognition, and medical image analysis. The ability of CNNs to extract local features and their ability to be invariant to translation, rotation, and scale make them a powerful tool in computer vision applications. Furthermore, CNNs can be used for tasks beyond image recognition, such as natural language processing and speech recognition.

In this research, deep CNNs are used to perform image recognition on images extracted from the dark web. The extracted images are categorized into 5 different categories of criminal activities, including drugs, weapons, bank cards, identity documents, and illegal currencies. The darkoob dataset is introduced as a valuable resource for training machine learning algorithms and deep learning models to improve the accuracy of detecting criminal activities in the dark web. The CNN algorithms used in this research are trained with artificially modified data to address the limitations of the dataset and to provide a more comprehensive dataset for detecting criminal activities.

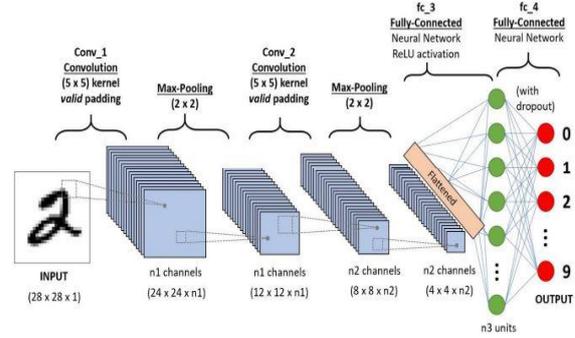

Figure. 8. An example of CNN neural network architecture

In the above figure, the input is a handwritten digit with a size of 1×28×28, where the height and width are 28 pixels, and there is one color channel. In color images, the channel size is 3, corresponding to the red, green, and blue channels. Multiple convolution filters or kernels are applied to the input image, where n1 is the number of filters. The terms filter and kernel are used interchangeably in the literature. In this case, the filter size is 5×5. The filter sizes must be the same within a layer, but they can be different between layers. The output size of the convolution layer in this case is 1×24×24, as valid padding is used with n1 filters. Valid padding refers to placing zeros around the image to maintain the output size the same as the input size. Then, a max pooling layer is applied to reduce the size of the feature maps and extract more complex features between points. The pooling operation is important as it reduces computational costs and increases the training speed of CNNs. Next, another convolutional layer and max pooling layer are applied. Finally, the resulting feature map is connected to a fully connected layer, which is connected to a softmax layer for classification. These are the main layers used in CNNs.

In this article, a neural network was designed using CNNs, where the architecture is shown in Figure (11). The network was trained using a custom dataset that was collected, and its performance is reported in the accuracy and error graphs of the network.

The network architecture consists of multiple layers of convolution and pooling, followed by a fully connected layer and a softmax layer for classification. The convolution layers apply multiple filters to the input image to extract local features, which are then down sampled by the pooling layers to reduce the dimensionality of the feature maps. The fully connected layer combines the extracted features and learns the weights to classify the input image into the desired categories.

The network was trained using backpropagation and stochastic gradient descent to minimize the categorical cross-entropy loss function. The training process was repeated for multiple epochs until convergence.

The results show that the network achieved high accuracy in classifying the images into their respective categories. The use of CNNs allowed for the extraction of complex features from the input images, which improved the accuracy of the classification task. The performance of the network was evaluated using various metrics, such as precision, recall, and acuucray , which demonstrated its effectiveness in detecting criminal activities in the dark web.

In conclusion, CNNs are a powerful tool for image recognition tasks, particularly for detecting criminal activities in the dark web. The proposed network architecture and the darkoob dataset can contribute to the development of more effective solutions for combating criminal activities in the dark web and can be a valuable resource for law enforcement agencies and researchers in the field of cybersecurity.

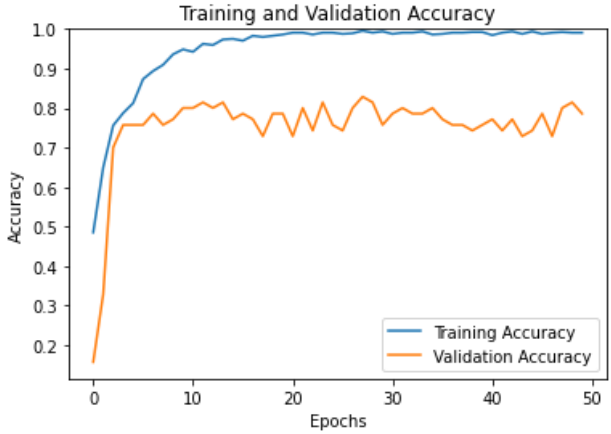

Figure. 9. Accuracy diagram of the architecture designed on the darkoob data set without data augmentation

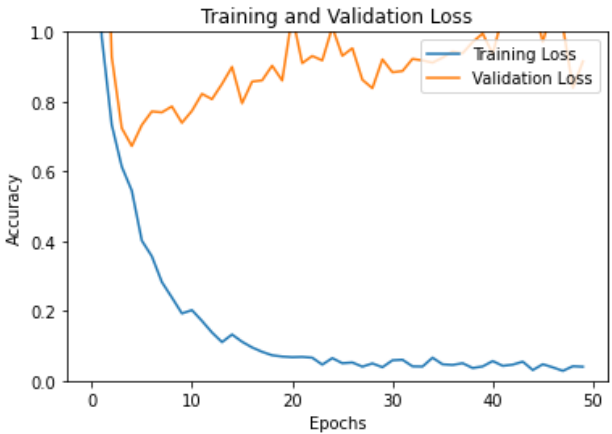

Figure. 10. Error diagram of the architecture designed on the darkoob data set without data augmentation

Since the amount of data for training a large neural network is scarce, we have used the transfer learning capabilities provided to us. Using the Resnet50 [14] network, which has been previously trained on the ImageNet dataset as a backbone, we add a few layers for classification operations at the end of the network. In other words, we fine-tune the network by training it with a specialized dataset. The results are reported in Figures 14 and 15, with an accuracy of 99.37% on the dataset.

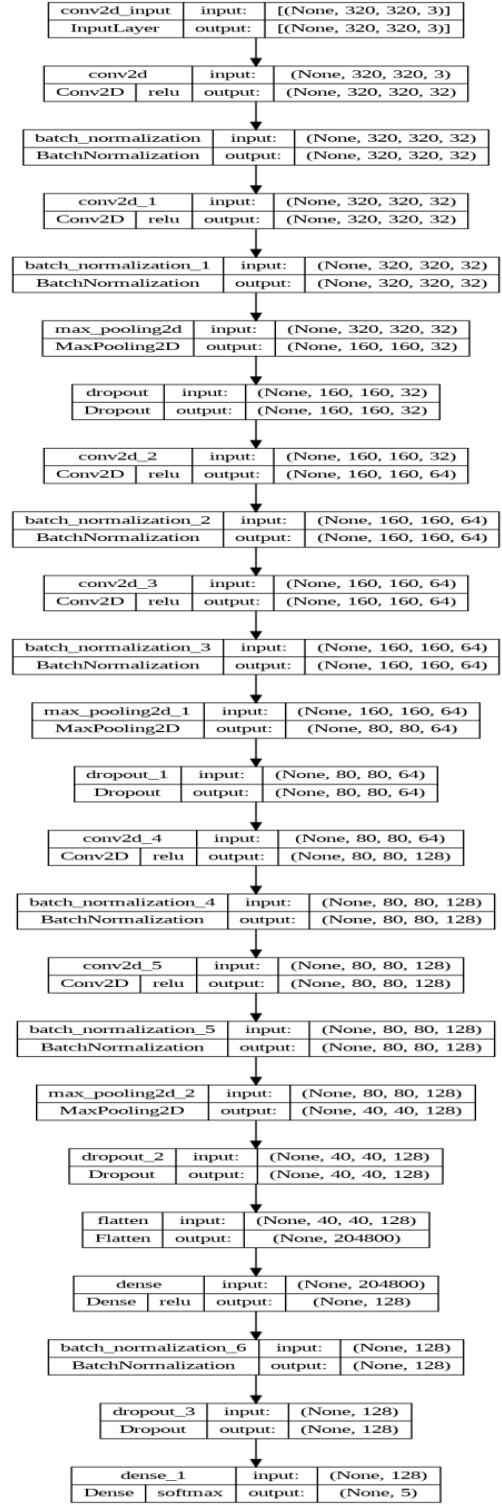

Figure. 11. designed network architecture

The number of parameters of the designed network is "26,504,485". After training for 50 epochs with the Categorical Cross Entropy loss function and SGD optimizer with a 0.001 learning rate and 0.9 momentum, the accuracy reaches 78.57%.

As the figure 10 shows, the network is overfitted on the training data. To solve this problem, we turn to transfer learning in the next section.

It should be noted that the results announced above are on the dataset without data augmentation.

After performing data augmentation and retraining the network on the augmented dataset, the results and chart are as follows:

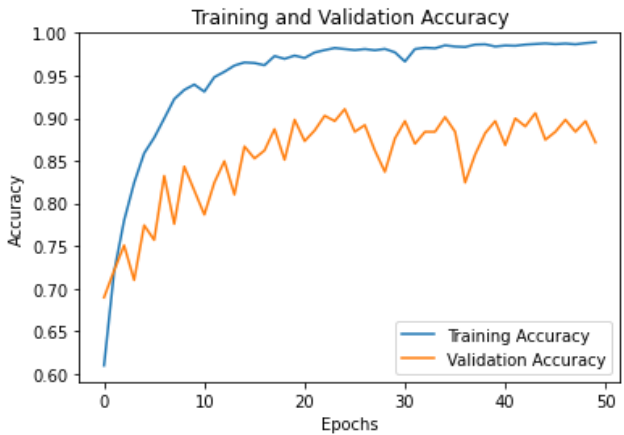

Figure. 12. Accuracy graph of the network designed on the darkoob data set by performing the data augmentation operation

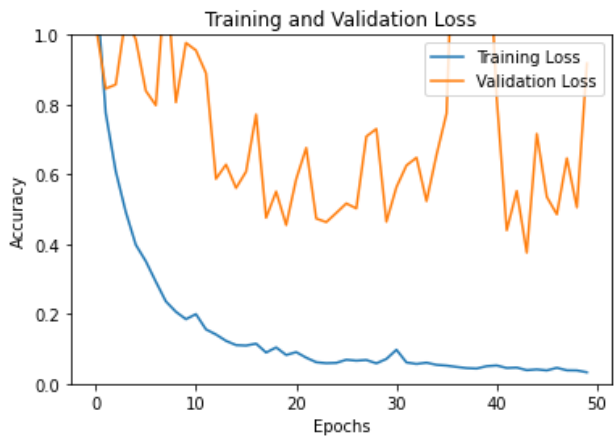

Figure. 13. Accuracy graph of the network designed on the darkoob data set by performing the data augmentation operation

As it is clear from the graph, the network was able to reach 87.15% accuracy, we go to transfer learning to increase the accuracy.

### 4.6. Transfer Learning

Finally, we turn to transfer learning on the augmented dataset. For transfer learning, the ResNet50 [14] network is used as the base network and feature extractor. At the end, a few fully connected layers are designed to perform image classification operations on 5 designed classes. The total number of parameters is around "26,343,301" million. After freezing the parameters of the base ResNet50 [14] network, the number of parameters for training reaches "2,755,589" million. Training the network with the Darkoob dataset achieves an accuracy of 99.37% on the validation dataset.

The designed network is trained for 20 epochs with the Categorical Cross Entropy loss function and SGD optimizer with a learning rate of 0.001 and momentum of 0.9. The error and accuracy charts using the transfer learning method are shown in Figures 14 and 15.

Ultimately, we employ transfer learning on the augmented dataset. For transfer learning, the ResNet50 [14] network is used as the base network and feature extractor. At the end, a few fully connected layers are designed to perform image classification operations on the 5 designed classes. The total number of parameters is around 26,343,301 million. After freezing the parameters of the base ResNet50 [14] network, the number of parameters for training reaches 2,755,589 million. Training the network with the darkoob dataset achieves an accuracy of 99.37% on the validation dataset.

The designed network is trained for 20 epochs with the Categorical Cross Entropy loss function and SGD optimizer with a learning rate of 0.001 and momentum of 0.9. The error and accuracy charts using the transfer learning method are shown in Figures 14 and 15.

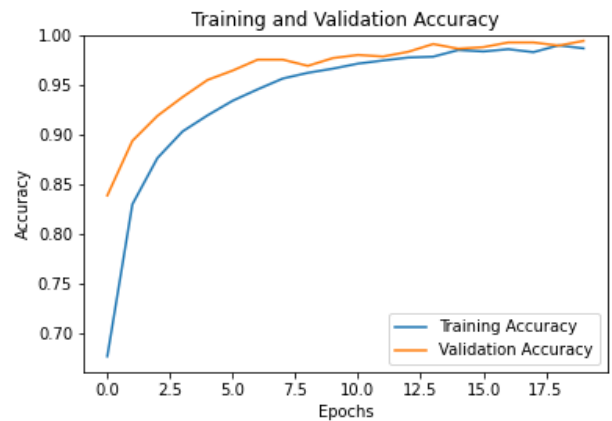

Figure. 14. The accuracy diagram of the network designed with transfer learning on the darkoob data set by performing the data augmentation operation.

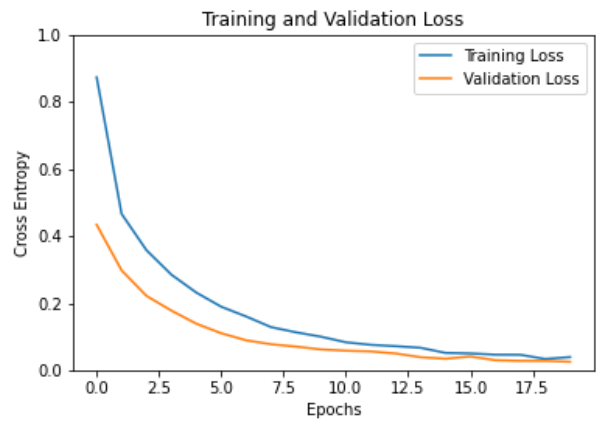

Figure. 15. The error diagram of the network designed with transfer learning on the darkoob data set by performing the data augmentation operation.

The network designed using transfer learning on the test data set has reached 94% accuracy for 100 data from different categories.

## 4.7. Keyword Extraction

KeyBERT[1] is an open-source Python package that is used for automatic keyword extraction. It utilizes the BERT model for text representation and extracts important keywords from the text using a novel method for generating hidden vectors.

KeyBERT employs a multi-head attention algorithm for selecting important text sentences and extracting keywords.

After extracting keywords using KeyBERT and converting each keyword into a feature vector , we measure the distance of each extracted keyword to the five predefined categories. Then, we assign each keyword to the category it is closest to, and consider it as a dark web activity.

In Figure 16, you can see the general steps for detecting the title of a dark web activity using keyword extraction:

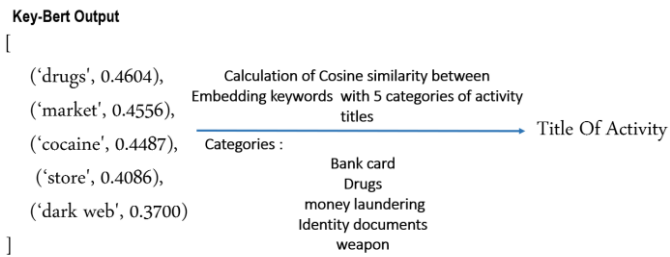

Figure. 16. The process of NLP Activity Detection

## 5. Evaluation Metric

Accuracy is one of the most important criteria in image classification, which we will explain in the next steps:

### 5.1. Precision & Recall

Accuracy is calculated by the number of positive predictions that are correctly predicted as positive. The measure of precision also calculates the ratio of true positives to all possible outcomes. These two measures have an inverse relationship with each other. Figure 17 shows the confusion matrix.

Figure. 17. confusion matrix

- The formula for calculating the precision standard is as follows:

$$Precision = \frac{TP}{TP+FP} \quad (1)$$

- The recall formula is calculated as follows:

$$Recall = \frac{TP}{TP+FN} \quad (2)$$

- And finally, the accuracy calculation formula is as follows:

$$Accuracy = \frac{(TP + TN)}{(TP + TN + FP + FN)} \quad (3)$$

In this research, we have reported the performance of our system based on the evaluation measure of accuracy, which is used to evaluate the performance of image classification algorithms.

## 6. Conclusion

Given the importance of detecting the title of activities in the dark web, which serves as a suitable place for criminals and smugglers to carry out illegal activities, this study aimed to introduce a custom dataset called darkoob for detecting the title of activities in 5 categories (drug trafficking, weapon trafficking, bank card trafficking, counterfeit currency trafficking, and identity document trafficking such as national ID cards and passports) in the dark web. Finally, a deep neural network CNN and NLP was designed and implemented for detecting the title of activities based on images and texts available on websites in the dark web, which achieved an accuracy of 94% in various scenarios.

---

[1] https://github.com/MaartenGr/KeyBERT

**Appendix**

In this implemented system, a web page has been designed that, by receiving .onion links and examining the content of the dark web using available images and text, identifies the activity title of that link.

Here are some examples of the system's outputs:

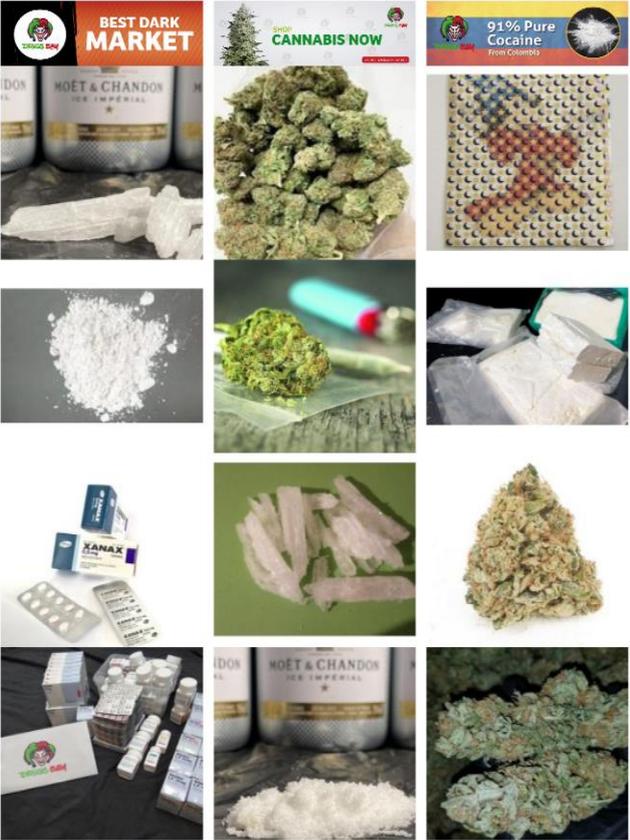
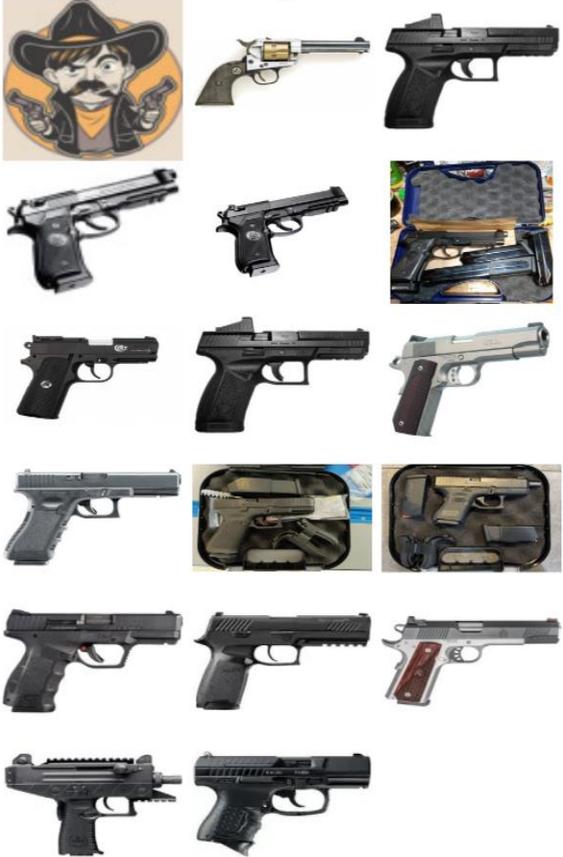

**References**

[1] da Cunha, B.R., MacCarron, P., Passold, J.F., dos Santos, L.W., Oliveira, K.A. and Gleeson, J.P., 2020. Assessing police topological efficiency in a major sting operation on the dark web. Scientific reports, 10(1), pp.1-10.

[2] Ebrahimi, M., Nunamaker Jr, J.F. and Chen, H., 2020. Semi-supervised cyber threat identification in dark net markets: a transductive and deep learning approach. Journal of Management Information Systems, 37(3), pp.694-722.

[3] Fernandez, E.F., Carofilis, R.A.V., Martino, F.J. and Medina, P.B., 2020. Classifying suspicious content in Tor Darknet. arXiv preprint arXiv:2005.10086.

[4] Fernandez, E.F., Carofilis, R.A.V., Martino, F.J. and Medina, P.B., 2020. Classifying suspicious content in Tor Darknet. arXiv preprint arXiv:2005.10086.

[5] Kamal, K.C., Yin, Z., Wu, M. and Wu, Z., 2019. Depthwise separable convolution architectures for plant disease classification. Computers and Electronics in Agriculture, 165, p.104948.

[6] He, S., He, Y. and Li, M., 2019, March. Classification of illegal activities on the dark web. In Proceedings of the 2nd International Conference on Information Science and Systems (pp. 73-78).

[7] Biswas, R., Fidalgo, E. and Alegre, E., 2017, December. Recognition of service domains on TOR dark net using perceptual hashing and image classification techniques. In 8th International Conference on Imaging for Crime Detection and Prevention (ICDP 2017) (pp. 7-12). IET.

[8] Krizhevsky, A., Sutskever, I. and Hinton, G.E., 2017. Imagenet classification with deep convolutional neural networks. Communications of the ACM, 60(6), pp.84-90.

[9] Tasci, T. and Kim, K., 2015. Imagenet classification with deep convolutional neural networks.

[10] Torrey, L. and Shavlik, J., 2010. Transfer learning. In Handbook of research on machine learning applications and trends: algorithms, methods, and techniques (pp. 242-264). IGI global.

[11] Fidalgo, E., Alegre, E., González-Castro, V. and Fernández-Robles, L., 2018. Illegal activity categorisation in DarkNet based on image classification using CREIC method. In International Joint Conference SOCO'17-CISIS'17-ICEUTE'17 León, Spain, September 6–8, 2017, Proceeding 12 (pp. 600-609). Springer International Publishing.

[12] Shorten, C. and Khoshgoftaar, T.M., 2019. A survey on image data augmentation for deep learning. Journal of big data, 6(1), pp.1-48.

[13] Albawi, S., Mohammed, T.A. and Al-Zawi, S., 2017, August. Understanding of a convolutional neural network. In 2017 international conference on engineering and technology (ICET) (pp. 1-6). Ieee.

[14] LeCun, Y., 1998. The MNIST database of handwritten digits. http://yann. lecun. com/exdb/mnist/.

[15] He, K., Zhang, X., Ren, S. and Sun, J., 2016. Deep residual learning for image recognition. In Proceedings of the IEEE conference on computer vision and pattern recognition (pp. 770-778).

[16] Object Management Group. Unified Modeling Language: Superstructure, Version 2.0, ptc/03-07-06, July 2003, http://www.omg.org/cgi-bin/doc?ptc/2003-08-02.